\documentstyle[11pt]{article}
\begin{document}
\begin{center}
  {\Large {\bf Topologically massive magnetic monopoles } }\\[8mm]
  {\large {\bf  A. N. Aliev, Y. Nutku and K. Sayg{\i}l{\i} } } \\[2mm]
    Feza Gursey Institute, P. O. Box 6 Cengelkoy
    81220 Istanbul, Turkey \\[2mm]
\end{center}
                                    
We show that in the Maxwell-Chern-Simons theory of topologically massive
electrodynamics the Dirac string of a monopole becomes a cone in
anti-de Sitter space with the opening angle of the cone determined by the
topological mass which in turn is related to the square root of the
cosmological constant. This proves to be an example of a physical system,
{\it a priory} completely unrelated to gravity, which nevertheless
requires curved spacetime for its very existence.
We extend this result to topologically massive
gravity coupled to topologically massive electrodynamics in the framework
of the theory of Deser, Jackiw and Templeton. The $2$-component spinor
formalism, which is a Newman-Penrose type of approach for three dimensions,
is extended to include both the electrodynamical and gravitational
topologically massive field equations. Using this formalism exact
solutions of the coupled Deser-Jackiw-Templeton and Maxwell-Chern-Simons
field equations for a topologically massive monopole are presented.
These are homogeneous spaces with conical deficit. Pure Einstein gravity
coupled to Maxwell-Chern-Simons field does not admit
such a monopole solution.

\section{Introduction}

  The principal result we shall present in this paper is a physical
system which at the outset is not related to gravity but which
nevertheless requires curved spacetime for its very existence.
This situation is best illustrated with the example of a magnetic monopole
in the framework of both electrodynamical and gravitational topologically
massive theories in $3$-dimensions. We find that the essential
new feature introduced by topological mass is to open up the Dirac string
of a monopole into a cone. The intuitive example of this phenomenon takes
place for Maxwell-Chern-Simons (MCS) theory
in a Riemannian $3$-manifold with Euclidean signature which shows that
solutions of the MCS field equations naturally lead us into de Sitter (dS)
space with conical deficit. Three dimensional flat spacetimes with,
or without conical deficit do not allow such a solution.

   In $4 n + 3$ dimensions there exists the Chern-Simons action
through which we can introduce topological mass into Maxwell's
electrodynamics and Einstein's gravity \cite{djt}. In the simplest
case of $n=0$ pure Einstein gravity has no propogating degrees of
freedom and no Newtonian limit \cite{djth}.
On the other hand, three dimensional gravity with
a pure Chern-Simons action is equivalent to the Yang-Mills gauge
theory of the conformal group and therefore is finite and exactly
solvable \cite{witten}. There is, however, a very interesting
non-trivial theory of gravitation in $2+1$ dimensions which has
been proposed by Deser, Jackiw and Templeton (DJT) \cite{djt}
where the gravitational Chern-Simons action is added
to the Hilbert action. This is the theory of topologically massive
gravity (TMG). It is a dynamical theory of gravity unique to three
dimensions and the geometry of its exact solutions is non-trivial.
Mathematically the DJT field equations pose an interesting challenge
in that they are qualitatively different from the Einstein field
equations while posessing their elegance and consistency. Clement
\cite{clement} has made the most thorough investigation of the
solutions of DJT field equations for TMG as well as TME which uncovered
many interesting effects due to topological mass. Self-dual solutions
of TME coupled to Einsteinian gravity were discussed by
Fernando and Mansouri \cite{mans} and by Dereli and Obukhov \cite{tekin}
who gave the general analysis.

This class of fields we shall consider falls outside
the domain of solutions considered earlier \cite{clement}-\cite{tekin}
and illustrate in its purest form some of the new
interesting effects that take place in the presence of topological mass.
Earlier we \cite{an} presented the spinor formulation of
TMG in terms of real $2$-component spinors which provides a very useful
formalism analogous to the Newman-Penrose formalism \cite{np} of general
relativity. We shall extend this formalism to include topologically
massive electrodynamics and gravity. This formalism is helpful for
constructing physically meaningful exact solutions of the coupled
DJT-MCS field equations. We shall use it to derive the exact solution
for a topologically massive magnetic monopole.

\section{Dirac Monopole}
\label{sec-dirac}

It will be useful to start our considerations with a brief review of the
Dirac monopole and its extension to TME in order to explain the essential
idea we shall use throughout this paper.
Maxwell's electrodynamics is given by the action principle
\begin{equation}
I_M = - \frac{1}{2}  \int  ( F - \frac{1}{2} d A ) \wedge ^*F
\label{maxwellaction}
\end{equation}
in Minkowski spacetime and leads to the Maxwell field equations
\begin{eqnarray}
F & =& d A, \quad  \Rightarrow   \quad d F  = 0 \label{bianchi}
\label{bianchi1} \\
d ^*F  & = &  0 \label{maxwelleq}
\end{eqnarray}
the first one of which is the second Bianchi identity. Dirac \cite{dirac}
pointed out that for a magnetic monopole eqs.(\ref{bianchi}) must fail at
least in one point on every Gaussian surface enclosing the monopole. The set
of all such points forms the Dirac string. The Maxwell potential that
satisfies these requirements is a generalization of the $1$-form obtained
for the polar angle $d \phi$ on the plane which is closed but not exact.
The $U(1)$ potential $1$-form and the field $2$-form for the Dirac
monopole are given by
\begin{eqnarray}
A & = & g \left[ 1-cos \theta  \right] d \phi,  \label{diracpot} \\
F & = & g \, \sin \theta \, d \theta \wedge d \phi, \label{diracfield}
\end{eqnarray}
the latter of which is the familiar element of area on $S^2$.
The semi-infinite Dirac string is at $\theta=0$ and the surface
integral
\begin{equation}
\int F = 4 \pi \, g
\label{mcharge}
\end{equation}
determines the monopole magnetic charge.

   We shall now consider the Euclidean Maxwell-Chern-Simons topologically
massive electrodynamics in order to illustrate the essential new idea
brought in by making the Dirac monopole topologically massive. With the
inclusion of the electromagnetic Chern-Simons term the action is given by
\begin{equation}
I_{MCS} = - \frac{1}{2} \int \left\{
( F - \frac{1}{2} d A ) \wedge ^*F - \nu\, d A \wedge A \right\}
\label{mcsaction}
\end{equation}
which yields the MCS field equation
\begin{equation}
d ^*F  = \nu  F
\label{mcseq}
\end{equation}
and the Bianchi identity (\ref{bianchi}) where $\nu$ is a coupling constant,
the electromagnetic topological mass. In order to satisfy these field
equations with a $U(1)$ potential $1$-form satisfying the properties of
the Dirac monopole (\ref{diracpot}) we must introduce a {\it deficit}
in the polar angle $\theta$ that led to the Dirac string. That is,
{\it topological mass has the effect of turning the string into a cone}.
Thus we should consider a field $2$-form of the type
\begin{equation}
F  =  g \, \sin (b\,\theta)  \, d \theta \wedge d \phi
\label{tmfield}
\end{equation}
where $b$ is a constant deficit parameter which will be related to
topological mass.
Now it is clear that the potential $1$-form (\ref{diracpot}) must be
modified but still lead to eq.(\ref{tmfield}) as the field.
This suggests that we consider a Riemannian manifold with the co-frame
consisting of a {\it modified} form of the left-invariant $1$-forms
of Bianchi Type $IX$ parametrized in terms of Euler angles
\begin{eqnarray}
\sigma^0 & = & d \psi + \cos (b\, \theta) \, d \phi \nonumber \\
\sigma^1 & = & - \sin (b \,\psi) \, d \theta +
              \cos (b \, \psi) \, \sin (b \,\theta) \, d \phi
              \label{cof} \\
\sigma^2 & = & \cos (b \, \psi) \, d \theta +
              \sin (b\, \psi) \, \sin (b \, \theta) \, d \phi \nonumber
\end{eqnarray}
satisfying the Maurer-Cartan equations
\begin{equation}
d\sigma^{i}=\frac{1}{2}\, C_{\;\;j k}^{i}\,\sigma^{j}
\,\wedge\sigma^{k}   \label{maurerc}
\end{equation}
with non-vanishing structure constants
\begin{equation}
 C_{\;\;12}^{0}=-C_{\;\;21}^{0}= b, \hspace{1cm}
 C_{\;\;20}^{1}=-C_{\;\;02}^{1}= b, \hspace{1cm}
 C_{\;\;01}^{2}=-C_{\;\;10}^{2}= b.
\label{struconst8}
\end{equation}
Then a $U(1)$ potential $1$-form of the type
\begin{equation}
A=- g \sigma^0
\label{potnew}
\end{equation}
will have all the desired properties and lead to the field $2$-form
(\ref{tmfield}). The clue to the satisfaction of the field equation
(\ref{mcseq}) for TME lies in the fact that
with the co-frame (\ref{cof}) the Cartan-Killing metric
$ d s^2  =  \eta_{ik} \, \sigma^i \otimes  \sigma^k$ with
$\eta_{ik}  =  diag.(1,1,1)$ becomes
\begin{equation}
d s^2  =   d \theta^2 + d \phi^2 + d \psi^2
  +  2  \cos (b \theta) \, d \psi \, d \phi
\label{ads3}
\end{equation}
which is simply de Sitter space with the polar angle suffering a defect.
The duality relations for the basis (\ref{cof}) immediately leads
to the result that for the potential $1$-form (\ref{potnew})
eqs.(\ref{mcseq}) of TME will be satisfied identically in dS provided
\begin{equation}
b=\nu,
\label{solbnu}
\end{equation}
the deficit in the Eulerian polar angle is identified with topological mass.
From the curvature of (\ref{ads3}) we find
\begin{equation}
\lambda= \frac{\nu^2}{4}
\label{cosmologicalconstant}
\end{equation}
relating the cosmological constant to electromagnetic topological mass.
For the case of Lorentzian signature, {\it c.f.} section \ref{sec-exact},
this would be anti-de Sitter spacetime.
The field $2$-form (\ref{tmfield}) is shown in figure 1  where because we
are in the Euclidean sector the deficit in the polar angle caused by
topological mass can be given an explicit illustration.
The maximal analytical extension of dS space is given in the chart
\begin{eqnarray}
\cos\left( \frac{\nu}{2} \theta \right)
\cos\left[ \frac{\nu}{2} (\phi + \psi) \right]
&=& \tanh (\nu \alpha ), \nonumber\\
\sin \left( \frac{\nu}{2} \theta \right) &=&
\frac{\sin(\nu \beta)}{\cosh (\nu \alpha)}, \label{tiris}\\
\frac{1}{2} (\phi - \psi) &=& \gamma \nonumber
\end{eqnarray}
whereby the metric (\ref{ads3}) becomes
\begin{equation}
d \tilde{s}^2 = \Omega^{-2} \left( d \alpha^2 + d \beta^2
+ \sin^2 (\nu \beta) \, d \gamma^2 \right)
\label{confmetric}
\end{equation}
with the conformal factor
$ \Omega = \frac{1}{2} \cosh (\nu \alpha) $. The incomplete Einstein static
cylinder is manifest in eq.(\ref{confmetric}).

  In the discussion of TME monopole we started out with MCS field equations
(\ref{bianchi1}) and (\ref{mcseq}) which are written in a general background.
The expectation was that these field equations will admit a solution
in flat background spacetime for a
physical system which is electrodynamic in nature and {\it a priory}
completely unrelated to gravity. This proved to be impossible.
With the missing cone in the field $2$-form (\ref{tmfield})
eqs.(\ref{bianchi1}) and (\ref{mcseq}) could only be satisfied
in curved space (\ref{ads3}) with a corresponding conical deficit.

Thus we arrive at a remarkable conclusion that a physical system
of electrodynamic type requires curved spacetime for its existence.

\section{Spinor formalism in $(2+1)$-dimensions}

    We shall now turn our attention to Lorentz signature and introduce
the Newman-Penrose version of TME equations. This type of study
for topologically massive gravity was given by Hall, Morgan and Perjes
\cite{hmp} and its $2$-component spinor description with differential
forms was constructed in \cite{an} which will henceforth be referred to
as ${\bf I}$. Here we shall extend this work by first presenting the
spinor formulation of TME and in section \ref{sec-sources} couple it
to TMG.

We begin by recalling some basic relations from ${\bf I}$. At each point
of a three dimensional space-time with the metric of Lorentz signature
we can introduce a pair of real 2-component spinors
\begin{equation}
\zeta^{A}_{\,(a)} =
\{ \, o^{A} \, ,  \;\; \iota^{A} \, \} \;\;\;\;\;\;\; A = 1, 2
\;\;\; a = 0, 1
\label{01}
\end{equation}
which will define the basis. Spinor indices will be raised and lowered by
the Levi-Civita symbol $\epsilon_{AB}$ from the right and the normalization
of the spin frame is given by $ o_A \iota^A  = 1 $ with all others vanishing
identically. It is evident that such a spin frame will imply the triad of
real basis vectors which can be connected to basis spinors through
the Infeld-van der Waerden symmetric quantities $ \sigma^{i}_{\;AB}$.
We recall that the co-frame is given by
\begin{equation}
\sigma_{i}^{\;AB} \, d x^i     =   \left(       \begin{array}{cc}
n  &  - \frac{1}{\sqrt{2}} m  \\ - \frac{1}{\sqrt{2}} m & l
\end{array}             \right)
\label{coframe}
\end{equation}
and the space-time metric is
\begin{equation}
 d s^{\;2} = l \otimes n + n \otimes l - m \otimes m
\label{metric}
\end{equation}
where $ l, n $ are null and $m$ is space-like. The Newman-Penrose intrinsic
derivative operators in the direction of the legs $l^i, n^i, m^i$ of the
triad are given by $D, \Delta, \delta$ respectively so that
\begin{equation}
 d = l \, \Delta + n \, D -  m \, \delta
\label{d}
\end{equation}
is resolution of the exterior derivative along the legs of the triad.
We recall that taking the exterior derivative of the basis 1-forms and
expressing the result in terms of the basis 2-forms yields the spin
coefficients through the solution of a linear algebraic system.
The result is given by eqs.({\bf I}. 23)
\begin{equation}                  \begin{array}{cll}
d \, l & = & - \epsilon \; l \wedge n +  (\alpha - \tau ) \;
           l \wedge m  - \kappa \; n \wedge m  \\[2mm]
d n & = &  \epsilon' \; l \wedge n  - \kappa' \; l \wedge m
- ( \alpha  + \tau' ) \; n \wedge m  \\[2mm]
d  m & = & ( \tau' - \tau ) \; l \wedge n - \sigma' \;
l  \wedge m  - \sigma \; n  \wedge m
\end{array}
\label{dlnm}
\end{equation}
from which the spin coefficients are obtained through the solution of
a linear algebraic system. Here prime denotes the symmetry operation
resulting from the interchange of $l$ and $n$ leaving $m$ fixed.
We note that $ \alpha' = - \alpha $.

Earlier \cite{an} we had not introduced the spinor equivalent of
the basis 2-forms which are wedge products of the Infeld-van der
Waerden matrices of basis 1-forms (\ref{coframe})
\begin{equation}
\label{wedgep}
L^{AXBY} = \sigma^{AX}\wedge \sigma ^{BY}
\end{equation}
with the spinor equivalent
\begin{equation}
\label{2fspdec}
L^{AXBY}  = L^{AB} \, \epsilon^{\;XY}
+ L^{XY}\,\epsilon^{\;AB} 
\end{equation}
due to skew symmetry in the pair of indices $ AX $ and  $ BY $
and the basic spinor relation ({\bf I}. 11).
The 2-component spinors $\, L^{AB}\,$ and $\,L^{XY} \,$  are real
and symmetric $\, L^{AB} = L^{(AB)} \,$
$\, L^{XY} = L^{(XY)} \,$ and we have
\begin{eqnarray}
L^{00} & = & -\frac{1}{\sqrt{2}}\,n \wedge m \nonumber \\
L^{01} & = & -\frac{1}{2}\,l \wedge n  \label{2fbasis} \\
L^{11} & = & \frac{1}{\sqrt{2}}\,l \wedge m \nonumber 
\end{eqnarray}
for the basis $2$-forms.
Using the definition of Hodge star operator ({\bf I}. 57) and the
completeness relation ({\bf I}. 8) we find that
\begin{equation}                  
^*{ \hspace{-0.5mm} } \left(  l  \wedge n  \right)  =  - m \;,\;\;\;\;\;\;
^*{ \hspace{-0.5mm} } \left(  l  \wedge m  \right)  =  - l \;,\;\;\;\;\;\;
^*{ \hspace{-0.5mm} } \left(  n \wedge  m  \right)  =  n
\label{hodge}
\end{equation}
determine the duals of the basis 2-forms in (2+1)-dimensions. All other
duality relations can be obtained from $^{**}=1$.

\section{Topologically massive electrodynamics}

  In general relativity the spinor approach has turned out to be very useful
for the investigation of physically interesting solutions of the Einstein
and Maxwell equations. This should be the case for three dimensional
spacetimes as well. So we shall now derive the TME equations in the
spinor formalism. We recall that the field $2$-form is given by
\begin{equation}
\label{max2form}
F=\frac 12\,F_{i k }\,dx^i \wedge dx^k = \frac
12\,F_{AXBY} \,\sigma^{AX} \wedge \sigma^{BY} 
\end{equation}
in terms of the basis 2-forms (\ref{2fbasis}). Using the same considerations
that led to (\ref{2fspdec}) we decompose the electromagnetic spinor
\begin{equation}
\label{emspdec}
F_{AXBY} =
\varphi_{AB}\,\epsilon_{XY} + \varphi_{XY}\,\epsilon_{AB} 
\end{equation}
where $\varphi_{AB}$ and $\varphi_{XY}$ are symmetric
second rank real 2-spinors. Taking into account this relation in
(\ref{max2form}) together with (\ref{2fspdec}) and (\ref{2fbasis})
we find that
\begin{equation}
\label{max2formexp}
 F = \varphi _0\, m\wedge n - \varphi _1\,l\wedge n + \varphi _2\,l\wedge m  
\end{equation}
where
\begin{eqnarray}
\label{maxsc}
\varphi _0 : & = & \sqrt{2}\,\varphi_{00} = F_{i k } l^i m^k
\nonumber \\
\varphi _1 : & = & 2\,\varphi_{01} = F_{i k } l^i n^k       \\
\varphi _2 : & = & \sqrt{2}\,\varphi_{11} = F_{i k } m^i n^k   .
\nonumber
\end{eqnarray}
are the three real triad scalars of the electromagnetic field.
Under the action of the prime operation the
triad scalars undergo the transformation
$\varphi _1  \leftrightarrow - \varphi_1, \; \;
\varphi _0 \; \leftrightarrow \; -  \varphi_2$.
We also need the dual of the field 2-form (\ref{max2formexp}) which
is given by
\begin{equation}
\label{maxdual}
^*{ \hspace{-0.5mm} } F = -\varphi _0\, n + \varphi _1\,m - \varphi _2\,l  
\end{equation}
due to the relations (\ref{hodge}).
Using eqs.(\ref{max2formexp}) and (\ref{maxdual}) together with
(\ref{d}) and  (\ref{dlnm}) the field eqs.(\ref{mcseq})
assume the form
\begin{eqnarray}
(\delta -\alpha -\tau'-\nu )\,\varphi _0 - (D -\sigma )\,\varphi _1
- \kappa\,\varphi _2 & = & 0, \nonumber\\
(\Delta -\sigma' )\,\varphi _1 - (\delta +\alpha -\tau +\nu )\,\varphi _2
+ \kappa'\,\varphi _0 & = & 0,     \label{sptme}   \\
(\Delta +\epsilon' )\,\varphi _0 - (D +\epsilon )\,\varphi _2
-(\tau'-\tau +\nu )\,\varphi _1 & = & 0 \nonumber
\end{eqnarray}
and the Bianchi identity (\ref{bianchi}) is given by
\begin{equation}
\label{spbianchi}
(\delta -\tau -\tau')\,\varphi _1 - (D +\epsilon -\sigma)\,\varphi _2
-(\Delta +\epsilon'-\sigma')\,\varphi _0 = 0.
\end{equation}
These are the Newman-Penrose version of TME equations. We note
that under the prime operation the first two equations in (\ref{sptme})
go over into each other, provided that the sign of the TME coupling
constant is also changed simultaneously, whereas the last equation in
(\ref{sptme}), as well as the Bianchi identity (\ref{spbianchi}) remain
invariant.

   Finally, we note that the Maxwell stress tensor
\begin{equation}
T_{ik} = - F_{i}^{\;\;j} F_{kj} + \frac{1}{4}\,g_{ik}\,F^{mn} F_{mn}
\label{maxwellstress}
\end{equation}
can be expressed in terms of the electromagnetic triad scalars
as follows
\begin{eqnarray}
T_{ik}=\varphi_{0}^2\,n_{i} n_{k} + \varphi_{1}^2\,l_{(i} n_{k)}
+ \varphi_{2}^2\,l_{i} l_{k}+
(\varphi _0\,\varphi _2 +  \frac{1}{2}\,\varphi_1)\,m_i m_k
\nonumber \\
- 2\, \varphi_{0} \varphi_{1}\,n_{(i} m_{k)}
- 2\, \varphi_{1} \varphi_{2}\,l_{(i} m_{k)}
\label{trmaxstress}
\end{eqnarray}
where round parantheses denote symmetization. The Chern-Simons term
makes no contribution to the energy-momentum tensor.

\section{Topologically massive gravity with sources}
\label{sec-sources}

Deser, Jackiw and Templeton's \cite{djt} theory of topologically
massive gravity overcomes the dynamically trivial character of
Einsteinian gravity in three dimensions.
In our previous work {\bf I} we wrote the DJT field equations
in terms of differential forms with triad scalar coefficients.
Here we shall extend this formalism to topologically massive
gravity with sources, in particular TME. It is a property of $3$ dimensions
that symmetric second rank tensors can be written as $2 \times 2$ matrices
of $2$-forms which enables us to write the field equations in compact form.
For this purpose we shall start by constructing the energy-momentum $2$-form
in three dimensions.

\subsection{Energy-momentum 2-form}

The spinor equivalent of the symmetric energy-momentum tensor $ T_{ik} $
admits the decomposition
\begin{equation}
\label{spemt}
T_{ABXY}= \frac{1}{2}\,\left(T_{ABXY}+T_{ABYX} \right) +
\frac{1}{2}\,\left(T_{BAYX} - T_{ABYX} \right)
\end{equation}
where the first paranthesis is symmetric in pair of indices $A$, $B$ and
$X$, $Y$, while the second one is skew in the same pair of indices. Applying
the basic spinor relation ({\bf I}. 11) to the second group of indices
we obtain
\begin{equation}
\label{spemt1}
T_{AXBY}= S_{ABXY} + \frac{1}{3}\,T\,\epsilon_{AB}\,\epsilon_{XY}
\end{equation}
where
$$ T= T_{AX}^{\;\;AX}= T_i^{\;i} $$
and
$$ S_{ABXY}= T_{(AB)XY} = T_{AB(XY)}=T_{(AB)(XY)} $$
is the trace-free part of the energy-momentum tensor. Next, we shall use
the trace-free part of the energy-momentum tensor to construct the spinor
valued energy-momentum 2-form
\begin{equation}
T_{A}^{\;\;B}= S_{A\;\;\;XY}^{\;B}\,\Sigma^{XM} \wedge \Sigma_{M}^{\;\;Y}
\label{em2form}
\end{equation}
where
\begin{equation}
\Sigma_{A}^{\;\;B}=\frac{1}{\sqrt{2}}\,\,\sigma_{A\;\;i}^{\;\;B} \,\,dx^i .
\label{sigmaf}
\end{equation}
are obtained by by lowering a spinor index in eqs.(\ref{coframe}).
We can now express the components of the energy-momentum 2-form
in terms of triad scalars. We have
\begin{eqnarray}
 2 T_{0}^{\;\;0}  &    = &
 \left( T_{02} + \frac{1}{2}\,T \right)\, l \wedge n
- T_{12}\, l \wedge m + T_{01}\, n \wedge m \nonumber \\
\sqrt{2}\,T_{0}^{\;\;1} & = & - T_{01}\, l \wedge n  +
\frac{1}{2}\, T_{02}\,l \wedge m - T_{00}\, n \wedge m
\label{compemt2}                \\
\sqrt{2}\, T_{1}^{\;\;0} & = & T_{12}\, l \wedge n - T_{22}\, l \wedge m
 + \frac{1}{2} \, T_{02}\, n \wedge m \nonumber
 \end{eqnarray}
where we have introduced the definitions
\begin{equation}                       \begin{array}{llllllllll}
 T_{00} & := & S_{0000}, \hspace{5mm}      &
 T_{01} & := & \sqrt{2}\, S_{0010},  \hspace{5mm}        &
 T_{02} & := & 2\, S_{0011},  \hspace{5mm}           \\ [2mm]
 T_{11} & := & S_{0101},      &
 T_{12} & := & \sqrt{2}\,S_{0111},      &
 T_{22} & := & S_{1111}
 \end{array}
\end{equation}
which consist of the triad scalars
\begin{equation}                  \begin{array}{lllllllll}
T_{00}  & = & T_{ik} \, l^i l^k ,     &
T_{01}  & = & T_{ik} \, l^i m^k  ,    &
T_{02}  & = & T_{ik} \, m^i m^k   ,    \\[2mm]
T_{11}  & = & T_{ik} \, l^i n^k - \frac{\textstyle{1}}{\textstyle{3}} \,T , &
T_{12}  & = &  T_{ik} \, n^i m^k ,  &
T_{22}  & = & T_{ik} \, n^i n^k
\end{array}
\label{ttcoeff}
\end{equation}
of the energy-momentum tensor. We note that under the prime operation the
index 1 remains unchanged while $ 0 \leftrightarrow 2 $.

\subsection{DJT field equations with sources}

The DJT field equations of TMG with sources are given by
\begin{equation}
G^{ik} + \frac{1}{\mu}\, C^{ik} = \lambda \, g^{ik} - \mbox\ae \,T^{ik}
\label{djteq}
\end{equation}
where $ G^{ik} $ is the Einstein tensor and $C^{ik}$
is Cotton's conformal tensor of three-dimensional manifolds. The constants
$\mu$ and $\mbox\ae$ are the DJT topological and Einstein matter coupling
constants with $\lambda$ standing for the cosmological constant.
The sign of the matter coupling constant is taken to be negative, in contrast
to four-dimensional gravity, to choose the physical
non ghost-like excitation modes. The above definition of the matrix of
energy-momentum 2-form along with the curvature and Cotton 2-forms
described by eqs.({\bf I}. 41) and ({\bf I}. 55) enable us to
write the field equations (\ref{djteq}) in the form
\begin{equation}
R_{A}^{\;\;B} + \frac{1}{\mu} \, C_{A}^{\;\;B} +
\left(\lambda - \frac{1}{2}\, \mbox\ae  T \right)\,
\Sigma_{A}^{\;\;M} \wedge \Sigma_{M}^{\;\;B}= \mbox\ae T_{A}^{\;\;B}
\label{djteq2}
\end{equation}
that consist of matrices of differential $2$-forms with triad scalar entries.
The expression for the DJT field equations follows from the substitution
of the results in eqs.(\ref{compemt2}), ({\bf I}. 42) and ({\bf I}. 60-62)
into eq.(\ref{djteq2}). Thus we arrive at the following set
of DJT field equations
\begin{eqnarray}
 D \Phi_{12}  - \Delta \Phi_{10}  - 3 ( \tau - \tau' ) \Phi_{11}
- \epsilon' \Phi_{01} +  \epsilon \Phi_{12}  
+ \kappa \Phi_{22}- \kappa' \Phi_{00}  \nonumber  \\ 
=   \mu \,\Phi_{02} - \frac{1}{2}\,\mu ( \lambda
+ 9\,\Lambda + \mbox\ae T_{02})         \label{tmg1}
\end{eqnarray}
\vspace{0.5mm}
\begin{eqnarray}
\delta \Phi_{01} - D \Phi_{02}  - 2 \kappa  \Phi_{12}
- (  \alpha + 2 \tau' ) \Phi_{01}
 + \sigma' \Phi_{00}         
+3 \sigma \Phi_{11}+\frac{9}{4}D\Lambda     \nonumber \\
  =    - \mu (\Phi_{01}- \frac{\mbox\ae}{2}\,T_{01}) \nonumber \\
 \delta \Phi_{12} - \Delta \Phi_{02}  - 2 \kappa'  \Phi_{01}
 + (  \alpha - 2 \tau ) \Phi_{12}   
+ \sigma \Phi_{22}  
 +3 \sigma' \Phi_{11}
 +\frac{9}{4}\Delta\Lambda  \label{tmg2}  \\
 =   \mu (\Phi_{12} - \frac{\mbox\ae}{2}\,T_{12}) \nonumber
\end{eqnarray}
\vspace{0.5mm}
\begin{eqnarray}   
 D \Phi_{11} -  \Delta \Phi_{00}
 + ( \tau' -  2 \tau ) \Phi_{01}
 - 2  \epsilon' \Phi_{00}  +
  \kappa \Phi_{12}-\frac{3}{4} D\Lambda  \nonumber \\
 =   \mu (\Phi_{01}- \frac{\mbox\ae}{2}\,T_{01})     \nonumber \\
\Delta \Phi_{11} -  D \Phi_{22} + ( \tau -  2 \tau' ) \Phi_{12 }
- 2 \epsilon \Phi_{22} 
 + \kappa' \Phi_{01} -\frac{3}{4}\Delta\Lambda  \label{tmg3}  \\
 =   - \mu (\Phi_{12} - \frac{\mbox\ae}{2}\,T_{12}) \nonumber
\end{eqnarray}
\vspace{0.5mm}
\begin{eqnarray}   
 \Delta \Phi_{01} - \delta \Phi_{11} + \kappa' \Phi_{00}  + 3 \tau \Phi_{11}
 - \sigma \Phi_{12}  
 + ( \epsilon' - \sigma' )  \Phi_{01} + \frac{3}{4}
\delta\Lambda         \nonumber    \\ 
 =   -\frac{1}{2}\, \mu \,\left[( \lambda  +  \Phi_{02} )
 - \mbox\ae (T_{11} +\frac{1}{3}\,T) \right]  \nonumber    \\
D \Phi_{12} - \delta \Phi_{11} + \kappa \Phi_{22}  + 3 \tau' \Phi_{11}
 - \sigma' \Phi_{01}            
 + ( \epsilon - \sigma )  \Phi_{12}
 +\frac{3}{4}\delta\Lambda       \label{tmg4}      \\ 
 =   \frac{1}{2} \mu \left[( \lambda  +  \Phi_{02} ) -
 \mbox\ae (T_{11} +\frac{1}{3}\,T)\right]    \nonumber
\end{eqnarray}
\vspace{0.5mm}
\begin{eqnarray} 
  D \Phi_{01}  -  \delta \Phi_{00}
  +3 \kappa  \Phi_{11}  - ( \epsilon + 2 \sigma ) \Phi_{01} 
+ ( \tau' + 2 \alpha )  \Phi_{00}  \nonumber \\
  =    \mu \,(\Phi_{00} - \frac{\mbox\ae}{2}\, T_{00}) \nonumber \\
\Delta \Phi_{12}  -  \delta \Phi_{22}
  +3 \kappa'  \Phi_{11}  - ( \epsilon' + 2 \sigma' ) \Phi_{12}
 + (  \tau - 2 \alpha )  \Phi_{22}     \label{tmg42}     \\
  =   - \mu \,(\Phi_{22} - \frac{\mbox\ae}{2}\,T_{22}).  \nonumber
\end{eqnarray}
Finally the DJT field equations (\ref{djteq}) imply the trace relation
\begin{equation}
\lambda + \frac{1}{6} R =  \frac{1}{3}\,\mbox\ae T
\label{trace}
\end{equation}
since the Cotton tensor is traceless.
We note that the first equation in the above set remains
invariant and the rest equations in each pair go into other
under the prime operation provided that the sign of the DJT coupling
constant is also changed simultaneously.

\section{Exact solutions}
\label{sec-exact}

We shall now extend the discussion of the topologically massive Dirac
mono\-pole given in section \ref{sec-dirac} by presenting exact solutions
of the system of coupled DJT-MCS field equations that describe a
self-gravitating magnetic monopole.
For this purpose, we shall use the Newman-Penrose
version of the TME (\ref{sptme}) and TMG (\ref{tmg1}) -(\ref{tmg42})
field equations. We start with a homogeneous space which is
given by the left-invariant 1-forms
\begin{eqnarray}
\sigma^{0} & = & d\psi + \cosh{(b\,\theta)} \, d\phi \nonumber  \\
\sigma^{1} & = & - \sin{(b\, \psi)}\, d\theta
+  \cos{(b\, \psi)}\, \sinh{(b\,\theta)} \, d\phi \label{cof8} \\
\sigma^{2} & = & \cos{(b\,\psi)}\,d\theta +\sin{(b\,\psi)} \,
\sinh{(b\,\theta)} \, d\phi         \nonumber
\label{euler}
\end{eqnarray}
of {\it modified} Bianchi Type $ VIII $. There will be no confusion
as the earlier definition of $\sigma^i$ in eqs.(\ref{cof}) will not
be used in the rest of this paper.
The $1$-forms (\ref{cof8}) satisfy the
Maurer-Cartan equations (\ref{maurerc}) with structure constants
\begin{equation}
 C_{\;\;12}^{0}= - C_{\;\;21}^{0}= -b, \quad
 C_{\;\;20}^{1}= - C_{\;\;02}^{1}=b, \quad
 C_{\;\;01}^{2}= - C_{\;\;10}^{2}=b.
\label{struconst88}
\end{equation}
We define the co-frame
\begin{equation}
\omega^{0}=\lambda_{0}\,\sigma^{0}, \hspace{1cm}
\omega^{1}=\lambda_{1}\,\sigma^{1}, \hspace{1cm}
\omega^{2}=\lambda_{2}\,\sigma^{2},
\label{homframe}
\end{equation}
where $\lambda_{0}, \, \lambda_{1}$ and $\lambda_{2}$
are constant scale factors \cite{nb}.
The triad basis 1-forms will be defined by
\begin{equation}
l=\frac{1}{\sqrt{2}} \left( \omega^{0} - \omega^{1} \right) ,   \hspace{1cm}
n=\frac{1}{\sqrt{2}} \left( \omega^{0} + \omega^{1} \right) ,   \hspace{1cm}
m=\omega^{2}.
\label{null}
\end{equation}
Then the Ricci rotation coefficients
\begin{eqnarray}
\epsilon & = & \epsilon' = \sigma = \sigma' =0 , \nonumber
\label{hrotcoef}       \\
\tau & = & -\tau' =  -\frac{b\,\lambda_{2}}{2\,\lambda_{0} \lambda_{1}} ,
\nonumber \\
\kappa & = & -\kappa' = \frac{b}{2\,\lambda_{0}  \lambda_{1} \lambda_{2}}
\left(\lambda_{0}^{2} - \lambda_{1}^{2}\right)    ,          \\
\alpha & = & \frac{b}{2\,\lambda_{0}  \lambda_{1} \lambda_{2}}
\left( \lambda_{0}^{2} + \lambda_{1}^{2} -\lambda_{2}^{2} \right) \nonumber
\end{eqnarray}
are obtained by taking the exterior derivative of eqs.(\ref{null})
and comparing the result with eqs.(\ref{dlnm}).
The Ricci identities ({\bf I}. 45-49) now yield the expression for the scalar
of curvature
\begin{equation}
R = - \frac{b^2}{ 2 \lambda_{0}^2 \lambda_{1}^2 \lambda_{2}^2 }
(\lambda_{0}+\lambda_{1} +  \lambda_{2})
(\lambda_{1} +  \lambda_{2}  - \lambda_{0})
(\lambda_{0} - \lambda_{1} +  \lambda_{2})
(\lambda_{2} - \lambda_{0} - \lambda_{1})
\label{defR}
\end{equation}
which holds for both Bianchi Types $VIII$ and $IX$. It is
related to Menger curvature $K$ by
\begin{equation}
R \equiv \frac{b^2}{2} K^2
\label{defP}
\end{equation}
for three points on a space curve in a fictitious flat $3$-dimensional
Euclidean space where $ \lambda_{0}, \lambda_{1}$ and $\lambda_{2}$
denote the distances between these points \cite{menger}. If we consider
the limit $\lambda_i \rightarrow 0$ keeping one point fixed, then Menger
curvature reduces to the definition of curvature
in the Serre-Frenet formulas. This identification offers the possibility
of classifying homogeneous solutions of TMG in terms of
of space curves. Namely, for vacuum solutions \cite{nb} where
$\lambda_{0}=\lambda_{1}+\lambda_{2}$ the space curve is a straight line.

Now turning our attention to the potential 1-form, we note that we can take
\begin{equation}
A = - g \, \sigma^0
\label{potnewp}
\end{equation}
as in eq.(\ref{potnew}) but on {\it modified} Bianchi Type $VIII$
left-invariant $1$-forms (\ref{cof8}). Then the field $2$-form is given by
\begin{equation}
\label{max2mono}
 F = -\frac{1}{\sqrt{2}}\,\frac{g\,b}{\lambda_1\,\lambda_2}\,
 (l\wedge m - n\wedge m)
\end{equation}
and we get
\begin{equation}
\varphi_{0}= \varphi_{2} = -\frac{1}{\sqrt{2}}\,\frac{g\,b}
{\lambda_1\,\lambda_2} \;,  \hspace{1cm}
\varphi_{1}=0
\label{triadscmono}
\end{equation}
for the electromagnetric triad scalars.

Starting with the TMG frame (\ref{euler}), (\ref{homframe}) and
TME potential (\ref{potnewp})
all triad scalars reduce to constants and the field equations of
TME (\ref{sptme}) and TMG (\ref{tmg1}) -(\ref{tmg42}) consist of a set
of polynomials for constant scale factors $\lambda_i$
in terms of constants in the theory, namely the
topological masses $\mu, \nu$ and the cosmological constant $\mbox\ae$.

The TME eqs.(\ref{sptme}) are satisfied identically provided that
the deficit angle in eqs.(\ref{euler}) is determined by
\begin{equation}
b = - \nu \,\frac{\lambda_1 \lambda_2}{\lambda_0}
\label{tmesol}
\end{equation}
in terms of the electromagnetic topological mass. Next, we consider
the triad components of the Maxwell stress tensor using
eqs.(\ref{trmaxstress}) and (\ref{ttcoeff}) along with
eq.(\ref{tmesol}). For the nonvanishing components
of the energy-momentum tensor we have
\begin{equation}
T_{00} = T_{02} = T_{22}= 3\,T_{11} = - T = \frac{g^2\,\nu^2}
{2 \lambda_{0}^2}
\label{ttmono}
\end{equation}
which enter in the right-hand-side
of the DJT field equations (\ref{tmg1}) -(\ref{tmg42}).

\subsection{Two equal scale factors}

First we shall consider the case of vanishing cosmological
constant $\lambda=0 $ and show that it forces the equality
$\lambda_1=\lambda_2$ between scale factors.
The resulting solution is a generalization of the
Vuorio solution \cite{v}.
Writing the trace equation (\ref{trace}) in terms of
the Ricci rotation coefficients we have
\begin{equation}
  2\alpha \tau  +\kappa^{2}  -\tau^{2} = - \mbox\ae T
\label{sptrace}
\end{equation}
or using eqs.(\ref{hrotcoef}) we write it in the explicit form
\begin{equation}
\frac{1}{4} b^2 \, K^2 =  \mbox\ae T
\label{rlambda}
\end{equation}
and once again see that $K=0$ leads to vacuum solutions which is not of
interest in this paper. Henceforth we shall take $K\ne 0$.
The remaining DJT field equations reduce to
\begin{equation}
3 \kappa \Phi_{11} + ( 2 \alpha - \tau ) \Phi_{00} =
\mu \Phi_{00} +\frac{1}{2}\,\mbox\ae T
\label{heq1}
\end{equation}
\begin{equation}
3 \tau' \Phi_{11} + \kappa \Phi_{22} =
\frac{1}{2}\, \mu \Phi_{02}  ,
\label{heq2}
\end{equation}
and all other DJT field equations are identically satisfied.
The explicit form of these equations is obtained
by the substitution of the rotation coefficients (\ref{hrotcoef}) and
sources (\ref{ttmono}). When we compare the resulting expressions
eqs.(\ref{heq1}) and (\ref{heq2}) to
eq.(\ref{sptrace}) we arrive at a polynomial constraint
\begin{equation}
\frac{b^3}{\lambda_{0} \lambda_{1} \lambda_{2}} \,
\left(\frac{ \lambda_{1}^{2} + \lambda_{2}^2 - \lambda_{0}^2}
{\lambda_{2}^{2} - \lambda_{1}^2 - 3 \lambda_{0}^2} \right)
(\lambda_{2}^{2} - \lambda_{1}^{2}) K^2 = 0
\label{polm}
\end{equation}
which involves only the scale factors $\lambda_{i}$. We emphasize
that this constraint holds only in the case of vanishing
cosmological constant.
As it is readily seen from eq.(\ref{rlambda}) only the roots
$\lambda_{1}= \pm \lambda_{2}$  give rise to non-vacuum solutions
of the DJT field eqations. For the sake of certainty, we shall take
$\lambda_{1}= \lambda_{2}$, then the simultaneous solution
of eqs.(\ref{rlambda})- (\ref{heq2}) has the form
\begin{equation}
\lambda_{0}^2= -2 \mbox\ae g^2 \frac{\nu + 2 \mu}{2 \mu +3 \nu}
\;\;\;\;\;\;
\lambda_{1}^2 = \lambda_{2}^2 =
-2 \mbox\ae g^2  \frac{\mu+ \nu}{2 \mu +3 \nu}
\label{sol1}
\end{equation}
which is the generalization of the Vuorio solution \cite{v}
for a TME-TMG magnetic monopole.
Indeed, in the limiting case $g \rightarrow 0 $ when the monopole charge
vanishes, the denominator in eqs.(\ref{sol1})
vanishes as well, so that the ratio
\begin{equation}
\frac{2 \mbox\ae g^2}{2 \mu +3 \nu}= - \lambda_{1}^3 =
-\frac{\lambda_{0}^3}{8}
\label{ratio}
\end{equation}
is constant and the solution (\ref{sol1}) reduces to the Vuorio solution.

Finally, we note that the above solution of the DJT field
equations with a topologically massive monopole can be readily generalized
to include a cosmological constant. We find
\begin{equation}
\lambda_{0}^2= -\frac{2 \mbox\ae g^2 \nu^2}{2 \mu +3 \nu}\,
\frac{\nu + 2 \mu}{\nu^2 +4 \lambda}
\;\;\;\;\;\;
\lambda_{1}^2 = \lambda_{2}^2 =
-\frac{2 \mbox\ae g^2}{2 \mu +3 \nu}\,
\frac{\nu^3 + \mu \nu^2 - 4 \lambda \mu}{\nu^2 +4 \lambda}
\label{sol2}
\end{equation}
and the final metric is given by
\begin{eqnarray}
d s^2 & = &  -\frac{2 \mbox\ae g^2}{(2 \mu +3 \nu)(\nu^2 +4 \lambda)}
\left\{ \nu^2 ( 2 \mu + \nu )\, \left( \sigma^0 \right)^2
 \right. \nonumber \\[2mm]
&& \hspace{1cm} \left. - (\nu^3 + \mu \nu^2 - 4 \lambda \mu) \left[
 \left( \sigma^1 \right)^2 + \left( \sigma^2 \right)^2
 \right] \right\}
\label{final}
\end{eqnarray}
which generalizes the solution \cite{n} to the case
of a topologically massive magnetic monopole.

\subsection{Tri-axial solution}

When all three scale factors are different DJT-MCS field equations
admit a solution which is obtained by considerations very similar
to those given above. However, in this case polynomials involving
the scale factors are much more complicated. For the tri-axial
solution eq.(\ref{tmesol}) relating the deficit in the angle $\theta$
to TME mass $\nu$ remains the same.
But the result (\ref{cosmologicalconstant}) is now modified to
\begin{equation}
\frac{\lambda}{\nu^2 /4} =
(\lambda_{0}^2 -\lambda_{1}^2 -\lambda_{2}^2)^2 \,
 \frac{ \lambda_{1}^{\;2} \,\lambda_{2}^{\;2} \, K^2}{\lambda_{0}^{\;2} \, A}
\label{ln}
\end{equation}
where
\begin{equation}
A \equiv \lambda_{0}^{\;4} -3\lambda_{1}^{\;4} -3\lambda_{2}^{\;4}
+2\lambda_{0}^{\;2} \lambda_{1}^{\;2} +2\lambda_{0}^{\;2} \lambda_{2}^{\;2}
-2\lambda_{1}^{\;2} \lambda_{2}^{\;2}
\end{equation}
is another interesting polynomial.
The ratio of the two topological masses is given by
\begin{equation}
\frac{\mu}{\nu}  =  - \frac{A}{2 \lambda_{0}^{\;2}
(\lambda_{0}^{\;2} -\lambda_{1}^{\;2} -\lambda_{2}^{\;2})}
\label{munu}
\end{equation}
and finally 
\begin{equation}
g^2 \mbox\ae = - \frac{4}{A} (\lambda_{0}^{\;2} -\lambda_{1}^{\;2})
(\lambda_{0}^{\;2} -\lambda_{2}^{\;2})
\,  \lambda_{1}^{\;2} \, \lambda_{2}^{\;2} \, K^2
\label{ae}
\end{equation}
gives the relationship between the matter coupling constant and magnetic
charge to the scale factors in the metric.

\section{Conclusion}

We have presented the exact solution for a self-gravitating magnetic
mono\-pole in topologically massive gravity and electrodynamics.
Topological mass has the effect of turning the Dirac string into
a cone as well as imparting conical deficit to the homogeneous
Bianchi Type $VIII$ space. Just as in the case of topologically
massive electrodynamical monopole without gravity, we have a physical
system which requires curved spacetime for its existence. Furthermore,
the conical deficit due to the topologically massive field
can be accomodated only in curved spacetimes with a matching conical
deficit. This is evidently a general phenomenon
which we must expect when we consider monopole-type solutions of
coupled MCS and DJT field equations. It is interesting to note that
for pure Einstein gravity in $3$-dimensions coupled to Maxwell-Chern-Simons
field all three scale factors must coincide and there exists no
self-gravitating monopole solution.

\section{Acknowledgement}

We thank Stanley Deser for many interesting discussions.

\end{document}